\documentclass[10pt,a4paper]{article}
\usepackage{fullpage}
\usepackage{graphicx}
\usepackage{subfigure}
\usepackage{bm}
\usepackage{amsmath}
\usepackage{verbatim}
\usepackage{hyperref}

\def\bra#1{\mathinner{\langle{#1}|}}
\def\ket#1{\mathinner{|{#1}\rangle}}

\begin{document}

\title{Is there a dynamical cause of the spin-statistics connection?
}


\author{Aditya Gilra\footnote{Tata Institute of Fundamental Research, Mumbai, India. E-mail: agilra@tifr.res.in}}
\maketitle

\begin{abstract}
Extant proofs of the spin-statistics connection (SSC) are kinematical. C S Unnikrishnan has suggested that a dynamical interaction leading to the SSC would involve spin and perforce gravity, the only known universal force. For the scattering of two identical particles, he considers \cite{unni2004} the interaction of their spins with the gravito-magnetic field generated by their scattering motion through cosmic matter-energy. There the direct and particles-exchanged scattering amplitudes accumulate different quantum phases which provide the relevant bosonic/fermionic sign between them without applying the ad hoc SSC rule. Here it is shown that the scattering probabilities given by the standard implementation of SSC in quantum mechanics are actually not obtained from the above interaction for most initial spin states of the scattering particles. Instead, an unrealized peculiar dynamical interaction is required. Further, a spin-gravito-magnetic interaction as above (with caveats) would result in a large unmeasured spin-orbit coupling type effect on atomic energy levels. A comparison with a typical rotation based proof is also provided.
\end{abstract}

\section{Introduction}
\label{intro}

\paragraph*{}
The spin-statistics connection (SSC) is stated in quantum mechanics simply as: integer spin (using $\hbar=1$) particles follow Bose-Einstein statistics while half-odd-integer spin particles follow Fermi-Dirac statistics.

\paragraph*{}
Extant proofs of the spin-statistics connection are based on kinematics. Well established proofs of the spin-statistics theorem in quantum field theory rely on Lorentz invariance while most quantum mechanics based proposals rely on some form of rotational invariance or exchange implemented as rotation arguments. See \cite{duck1997} for a review.

\paragraph*{}
C S Unnikrishnan \cite{unni2004a} has pointed out that cosmic gravity is ever present and its effects must be taken into account. Further, it could provide a dynamical reason for the spin-statistics connection. A dynamical cause of the SSC must be via cosmic gravity \cite{unni2008} as gravity is the only known universal force and local mass effects are known to be too weak (cf. the Lense-Thirring effect due to earth) to cause the requisite phase accumulation.

\paragraph*{}
In \cite{unni2004}, he attempts to derive the SSC as a consequence of dynamics, specifically the coupling of moving particles having spin with the (approximately) homogeneous critical-density matter-energy distribution of the universe. This is outlined in section \ref{sec:unniproof}.

\paragraph*{}
In section \ref{sec:arbitrary}, I show that this proposal does not in fact give the standard quantum mechanical differential scattering cross-section, except for four special initial spin states of the scattering particles. In section \ref{sec:orbit}, the spin-gravito-magnetic coupling is applied (with caveats) to an electron in an atom leading to an unmeasured large fine structure splitting. The difference between the dynamical approach and rotation based approach to the SSC is discussed in section \ref{sec:rotation}. Section \ref{sec:conclusion} is the conclusion.

\section{Outline of dynamical approach to SSC in \cite{unni2004}}
\label{sec:unniproof}

\begin{figure}
\begin{center}
\subfigure[process 1: each particle turns ccw by $\varphi$]{\includegraphics[width=0.45\textwidth]{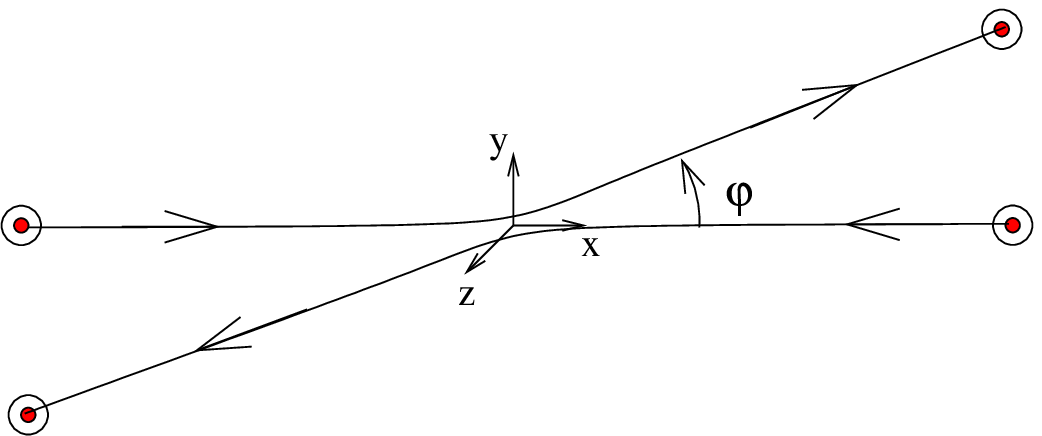}\label{fig:process1}}
~~
\subfigure[process 2: each particle turns cw by $\pi-\varphi$]{\includegraphics[width=0.45\textwidth]{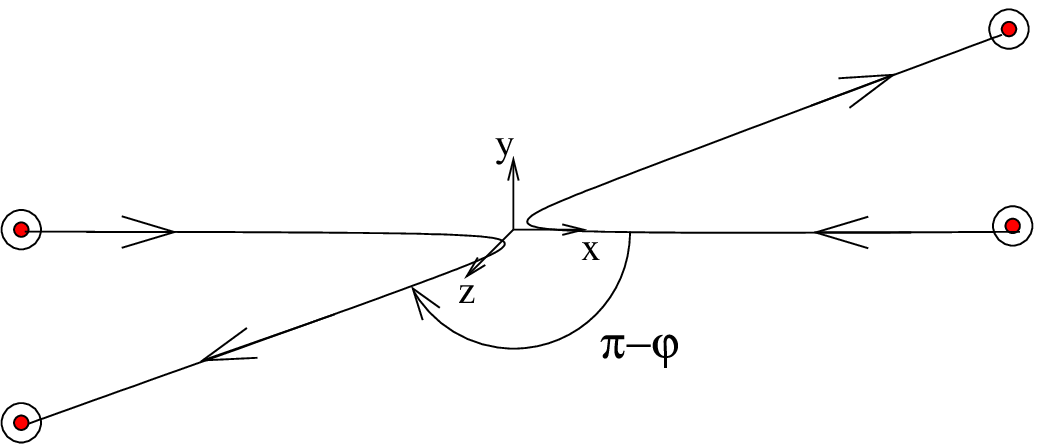}\label{fig:process2}}
\end{center}
\caption{Scattering of two identical particles into the same final state by two different processes (spins perpendicular to plane of scattering)}
\label{fig:qmech_scattering}
\end{figure}

\paragraph*{}
Consider the scattering of two identical spin $s$ particles, in the center of mass frame as shown in figure \ref{fig:qmech_scattering}, with both their spins aligned perpendicular to the plane of scattering. The interaction between the two particles is for clarity assumed spin-independent. Now the same final indistinguishable state can be attained by two processes as in figure \ref{fig:qmech_scattering}. The ad hoc SSC dictates that we must add/subtract the scattering amplitudes of these two processes depending on whether the spins are integer/half-odd integer and then square to find the transition probability / differential cross-section. However in the approach of \cite{unni2004}, the scattering amplitudes are consistently added while the relative sign appears as a dynamical phase difference between the two processes.

\paragraph*{}
In the direct scattering case, in the center of mass frame, the momenta of each particle rotates through angle $\varphi$ as in figure \ref{fig:process1}, while in the exchanged case, through angle $\pi-\varphi$ (oppositely) as in figure \ref{fig:process2}. The crucial assumption made in \cite{unni2004} is that this rotation of momenta is equivalent in physical effects to the matter-energy in the universe rotating oppositely by the same angle, while the particles forward scatter i.e. their momenta remain unchanged. The rotation of the matter-energy in the universe would generate a gravito-magnetic (Lense-Thirring) field $\bm{B}$ with the interaction Hamiltonian being $H_{int}=-\bm{\mu}_g.\bm{B}$, where the gravitomagnetic moment of a spin $\bm{s}$ particle is $\bm{\mu}_g=-\bm{s}$. 

\paragraph*{}
The gravitomagnetic field for a rotating cosmic matter-energy is now calculated.
The metric for a homogeneous isotropic (Robertson-Walker) universe with critical density in comoving coordinates $(t,x',y',z)$ is ($c=1$ throughout) 
\begin{equation}
ds^2=-dt^2+a^2(t)(dx'^2+dy'^2+dz^2).
\end{equation}
Transform to a frame $(t,x,y,z)$ rotating at angular velocity $\omega$ about the $z$ direction by applying the transformations:
\begin{equation}
x'=x\cos\omega t -y\sin\omega t~,~~~y'=x\sin\omega t +y\cos\omega t.
\end{equation}
These are not applied as a mere coordinate transformation. 
One is essentially measuring in a center of mass frame rotating at angular velocity $\omega$ or equivalently as if the whole universe is rotating oppositely.
The metric transforms to:
\begin{equation}
g_{\mu\nu}=\left(
\begin{array}{cccc}
 ~~-1+\left(x^2+y^2\right) \omega ^2 a^2(t) ~~&~~ -y \omega a^2(t) ~~ & ~~x \omega a^2(t)~~ & 0 \\
 -y \omega a^2(t) & a^2(t)~~~ & 0 & 0 \\
 x \omega a^2(t)& 0 & a^2(t) & 0 \\
 0 & 0 & 0 & a^2(t)
\end{array}
\right).
\end{equation}
 The off-diagonal components correspond to a gravito-magnetic vector potential
\begin{equation}
\mathbf{A}=(\frac{1}{2}y\omega,-\frac{1}{2}x\omega,0),
\end{equation}
setting $a(t_{now})=1$ for a critical universe. The case of non-critical RW universe ($k\ne0$) was not considered in \cite{unni2004}, but one can verify that the off-diagonal components still do not contain $k$. However, the freedom to set $a(t_{now})=1$ is not there.
The curl of $\mathbf{A}$ gives the gravito-magnetic field
\begin{equation}
\mathbf{B}=(0,0,-\omega).
\end{equation}
 
\paragraph*{}
Thus the interaction Hamiltonian is 
\begin{equation}
H_{int}=-\bm{s}.\bm{\omega}=-s_z\omega_z
\end{equation}
 which leads to a phase accumulation for the duration of scattering $t=0$ to $t=t_s$ on the wave-function of each particle:
\begin{equation}
\exp(-i\int_0^{t_s} H_{int} dt)=\exp(i\int_0^{t_s} s_z\omega_z dt) = \exp(i s_z\varphi).
\end{equation}
The Hamiltonian here is the generator of infinitesimal coordinate time translation, hence the use of coordinate time in the integral. The energy measured by a rotating observer will have a $\gamma$
factor \cite{mashhoon1988} which will cancel with the use of proper time in the integral yielding the same result.

\paragraph*{}
The phase on the post-scattering product wave-function of the two particles in the direct process would be $\exp(i2s\varphi)$ while in the exchanged process would be $\exp(i2s(-\pi+\varphi))$. These phase factors on the post-scattering product wave-functions cause a relative factor of $\exp(i2\pi s) = \pm 1$ depending on whether $s$ is integer / half-odd-integer between the two scattering amplitudes.

\paragraph*{}
Thus the correct relative sign between the scattering amplitudes is obtained as a quantum phase originating from cosmic gravity without recourse to an ad hoc rule.

\section{Generalization to arbitrary spins?}
\label{sec:arbitrary}

\paragraph*{}
The generalization of the above argument to spins in an arbitrary superposition of basis eigenkets was not done in \cite{unni2004}. Here it is shown that the differential cross-sections given by the standard implementation of SSC in quantum mechanics are not reproduced by the dynamical approach for most initial spin states of the particles. The standard implementation of SSC in quantum mechanics is to symmetrize (under particle label interchange) the wavefunction of identical integer spin particles and anti-symmetrize the wavefunction for identical half-odd-integer spin particles. Note that for two particles, there are only the possibilities of symmetrization and anti-symmetrization leading to bosonic and fermionic statistics respectively (See \cite{tino2000} for this and greater than two particles).

\begin{figure}
\begin{center}
\subfigure[process 1: each particle turns ccw by $\varphi$]{\includegraphics[width=0.45\textwidth]{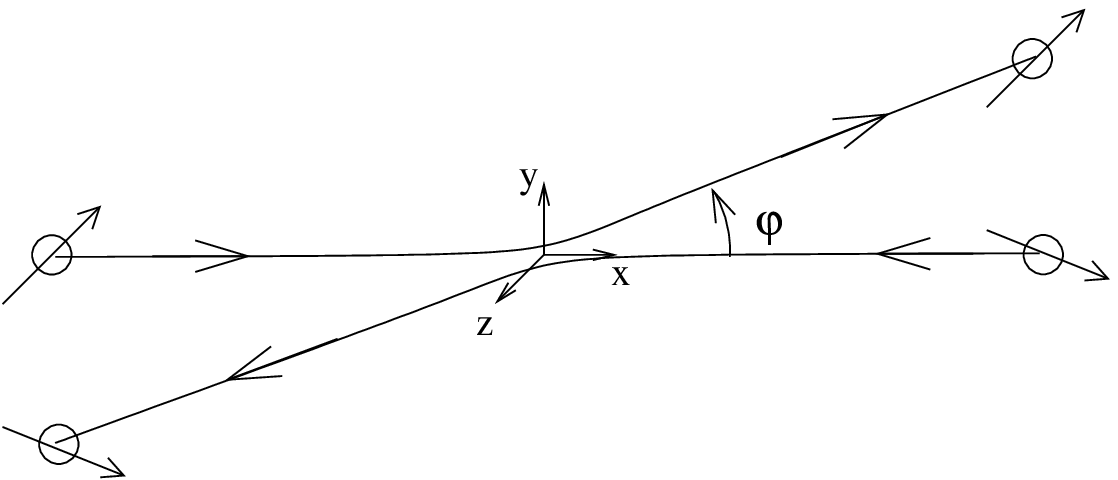}\label{fig:arbitrary_process1}}
~~
\subfigure[process 2: each particle turns cw by $\pi-\varphi$]{\includegraphics[width=0.45\textwidth]{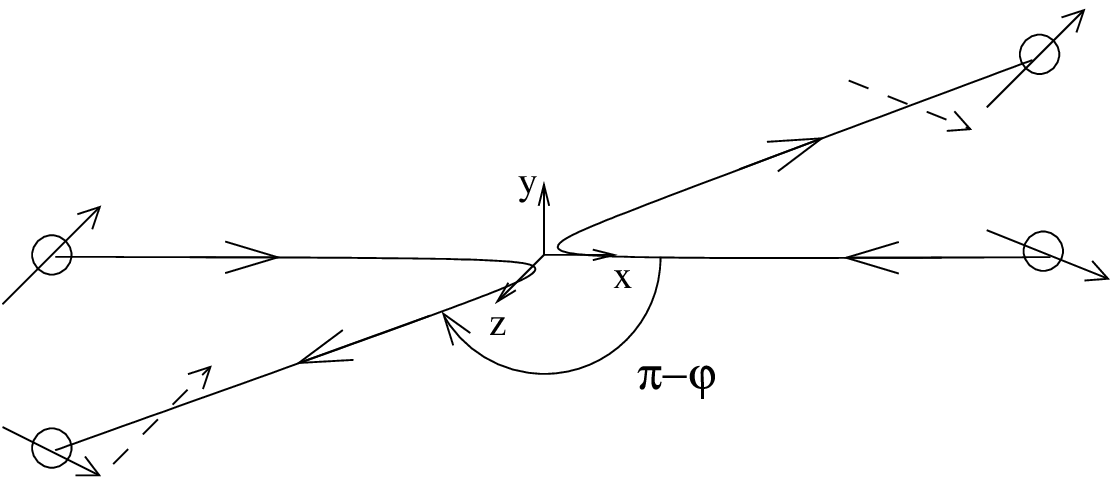}\label{fig:arbitrary_process2}}
\end{center}
\caption{Scattering of two identical particles (arbitrary spins) into the same final state by two different processes. Since the scattering is taken spin-independent, the spins in the particles-exchanged case retain their initial values (dashed lines) until they are measured in the final state (solid lines).}
\label{fig:arbitrary_qmech_scattering}
\end{figure}

\paragraph*{}
Consider center of mass frame scattering as in figure \ref{fig:arbitrary_qmech_scattering} with arbitrary initial spins. I follow an amalgamation of the formalisms in \cite{kessler1976} and \cite{taylor1972}. The initial asymptotic product state is:
\begin{equation}
\label{eqn:phi}
\ket{\phi}=\ket{\psi_1}\otimes\left(\sum_i\alpha_i\ket{i_s}\right)\otimes\ket{\psi_2}\otimes\left(\sum_j\beta_j\ket{j_s}\right).
\end{equation}
The left particle is in normalized initial state $\ket{\psi_1}\otimes(\sum_i\alpha_i\ket{i_s})$, where the kets $\ket{i_s}$ with $i=-s, \ldots, s$ in integer steps, are the orthonormal basis eigenkets of $s^2$ and $\hat{s}_z$, z-axis being perpendicular to the plane of scattering (x-y), and x-axis along initial momentum of left particle. Similarly for the right particle.

\paragraph*{}
The final direct and exchanged asymptotic states are respectively:
\begin{equation}
\label{eqn:phi'}
\ket{\phi'}=\ket{\psi_3}\otimes\left(\sum_i\alpha_i\ket{i_s}\right)\otimes\ket{\psi_4}\otimes\left(\sum_j\beta_j\ket{j_s}\right)
\end{equation}
and
\begin{equation}
\label{eqn:phi'_e}
\ket{\phi'_e}=\ket{\psi_4}\otimes\left(\sum_i\beta_i\ket{i_s}\right)\otimes\ket{\psi_3}\otimes\left(\sum_j\alpha_j\ket{j_s}\right).
\end{equation}
The first two slots in (\ref{eqn:phi}), (\ref{eqn:phi'}) and (\ref{eqn:phi'_e}) are for the space$\otimes$spin Hilbert space of the left particle while the last two slots are that for the right particle.
Note that $\ket{\phi'_e}$ is experimentally indistinguishable from $\ket{\phi'}$ and the final states shown in figures \ref{fig:arbitrary_process1} and \ref{fig:arbitrary_process2} corresponding to $\ket{\phi'}$ and $\ket{\phi'_e}$ respectively are exactly the same.

\paragraph*{}
The detection process is polarization sensitive measuring the final states $\ket{\psi_3}\otimes(\sum_j\alpha_j\ket{j_s})$ in the top detector and $\ket{\psi_4}\otimes(\sum_i\beta_i\ket{i_s})$ in the bottom detector, irrespective of which particle went where. One could as well choose to measure the more general final states $\ket{\psi_3}\otimes(\sum_i\gamma_i\ket{i_s})$ and $\ket{\psi_4}\otimes(\sum_j\delta_j\ket{j_s})$ in the top and bottom detectors respectively. But to keep expressions simple, we measure the final states that would occur for direct scattering.

\paragraph*{}
We consider the differential scattering cross-section from $\ket{\phi}$ to $\ket{\phi'}$. $\Lambda$ is a projection operator $\Lambda^2=\Lambda=\Lambda^\dagger$, which symmetrizes/anti-symmetrizes the wavefunction based on the spin of the scattering particles 
\begin{equation}
\label{eqn:Lambda}
\Lambda\ket{\phi'} = \frac{1}{2}(\ket{\phi'} \pm \ket{\phi'_e})
\end{equation} leading to a normalized state $\sqrt{2}\Lambda\ket{\phi'}$. The unusual numerical factors are required to maintain $\Lambda^2=\Lambda$. Thus the transition probability / differential cross-section is
\begin{align}
w &= \lvert 2\bra{\phi'}\Lambda^\dagger S\Lambda\ket{\phi}\rvert ^2    \\
					&= \lvert 2\bra{\phi'}\Lambda^\dagger S\ket{\phi} \rvert ^2, \label{eqn:symm_prob}
\end{align}
where $S$ is the scattering/evolution operator that takes the scattering potential/interaction into account and $\Lambda$ and $\Lambda^\dagger$ commute with it. Energy dependence is left out as it is irrelevant to the present discussion. For simplicity, interaction between particles is considered spin-independent so that $S=S_{space}\otimes 1$ leaves the spin part unchanged. Thus, the spins in the particles-exchanged case (figure \ref{fig:arbitrary_process2}) retain their initial values after scattering (dashed lines) until they are measured in the final state (solid lines).

\paragraph*{}
Using (\ref{eqn:phi}), (\ref{eqn:phi'}), (\ref{eqn:phi'_e}) and (\ref{eqn:Lambda}) in (\ref{eqn:symm_prob}) yields
\begin{equation}
\label{eqn:stdSSC}
w(\varphi)=\Biggl|f(\varphi)\pm f(-\pi+\varphi)\left(\sum_i\beta_i^*\alpha_i \sum_j\alpha_j^*\beta_j\right)\Biggr|^2
\end{equation}
where 
\begin{equation}
f(\varphi)=(\bra{\psi_3}\otimes\bra{\psi_4})S_{space}(\ket{\psi_1}\otimes\ket{\psi_2})
\end{equation}
and
\begin{equation}
f(-\pi+\varphi)=(\bra{\psi_4}\otimes\bra{\psi_3})S_{space}(\ket{\psi_1}\otimes\ket{\psi_2}).
\end{equation}

\paragraph*{}
For simplicity, the equations here and below are written only for scattering in the $x-y$ plane ($f$ not made a function of polar angle). There is no loss of generality as the axes can always be chosen in this manner.

\paragraph*{}
Now we consider the differential cross-section taking into account gravito-magnetic interaction in the manner of \cite{unni2004}. The evolution operator $S'$ now has an extra spin-dependent gravito-magnetic part over the $S$ above. Instead of symmetrizing/anti-symmetrizing, one merely adds the two scattering amplitudes, direct and exchanged, that yield the same final state $\ket{\phi'}$ and the correct sign should come from the spin-dependent gravito-magnetic part. The differential cross-section from $\ket{\phi}$ to $\ket{\phi'}$ (indistinguishable from $\ket{\phi'_e}$) is now
\begin{align}
w'(\varphi) = & |\bra{\phi'}S'\ket{\phi} + \bra{\phi'_e}S'\ket{\phi}|^2  \nonumber  \\
  = & \Biggl| f(\varphi)\left(\sum_j\alpha_j^*\alpha_j e^{i j \varphi}\sum_k\beta_k^*\beta_k e^{i k \varphi}\right)+   \nonumber \\
  & f(-\pi+\varphi)\left(\sum_l\beta_l^*\alpha_l e^{i l (-\pi+\varphi)}\sum_m\alpha_m^*\beta_m e^{i m (-\pi+\varphi)}\right) \Biggr| ^2. \label{eqn:dyn}
\end{align}

\paragraph*{}
Equation (\ref{eqn:dyn}) is not the same as (\ref{eqn:stdSSC}). In fact, except for the four cases when each initial spin has $s_z=+s$ or $s_z=-s$, there is an infinity of cases for which this dynamical argument fails.

\subsection{An example}
\label{subsec:example}

\paragraph*{}
If we take initially longitudinally polarized spin half ($s=1/2$) particles with both spins along $x$ axis ($\alpha_{-1/2}=1/\sqrt{2}$, $\alpha_{1/2}=1/\sqrt{2}$ and $\beta_{-1/2}=1/\sqrt{2}$, $\beta_{1/2}=1/\sqrt{2}$), the standard SSC gives
\begin{equation}
w = |f(\varphi) - f(-\pi+\varphi)|^2
\end{equation}
while the dynamical approach gives
\begin{align}
w' =& \Biggl| f(\varphi)\left(\dfrac{e^{-i\varphi/2}+e^{i\varphi/2}}{2}\right)^2
     + f(-\pi+\varphi)\left(\dfrac{e^{i(-\pi+\varphi)/2}+e^{-i(-\pi+\varphi)/2}}{2}\right)^2\Biggr|^2\\
     =& \bigl| f(\varphi)\cos^2(\varphi/2)+f(-\pi+\varphi)\sin^2(\varphi/2)\bigr|^2.
\end{align} 
In fact, at $\varphi=\pi/2$, we have for a spherically symmetric potential, $f(\varphi)=f(-\pi+\varphi)$ i.e. $f(\pi/2)=f(-\pi/2)$. Hence, $w=0$ while $w'=|f(\pi/2)|^2$. Thus, we note that the dynamical approach does not give the standard result even for identical initial spin states, except when they are perpendicular to the plane of scattering.

\paragraph*{}
For one spin along $x$ axis, the other along $-x$ axis, ($\alpha_{-1/2}=1/\sqrt{2}$, $\alpha_{1/2}=1/\sqrt{2}$ and $\beta_{-1/2}=-1/\sqrt{2}$, $\beta_{1/2}=1/\sqrt{2}$), standard SSC result is $w=|f(\varphi)|^2$, while dynamical approach gives 
\begin{equation}
w'=|(f(\varphi)-f(-\pi+\varphi))\cos^2(\varphi/2)|^2.
\end{equation}
At $\varphi=\pi/2$, we have $w=|f(\pi/2)|^2$, while $w'=0$.

\paragraph*{}
Thus the elastic scattering spin parameter 
\begin{equation}
C_{LL}=\frac{w(\rightarrow\rightarrow)-w(\rightarrow\leftarrow)}{w(\rightarrow\rightarrow)+w(\rightarrow\leftarrow)},
\end{equation}
(which measures the asymmetry in differential cross-section between initial spins longitudinally parallel and longitudinally anti-parallel), at $\varphi=\pi/2$ becomes $-1$ for standard SSC but $+1$ for dynamical approach.


\subsection{What would work?}

\paragraph*{}
Thus, we note that a dynamical interaction of the form $\bm{s}.\bm{\omega}=s_z\omega_z$ cannot reproduce the SSC. However, an interaction of the form
\begin{equation}
\label{eqn:works}
H_{int}=s\omega_z,
\end{equation} where $s$ is the spin of each of the identical particles (e.g. $1/2$ for electrons) and z-axis is perpendicular to the plane of scattering, would work for all spin states of the particles. With (\ref{eqn:works}), (\ref{eqn:dyn}) becomes 
\begin{align}
w'(\varphi) = & \Biggl| f(\varphi)\left(\sum_j\alpha_j^*\alpha_j e^{i s \varphi}\sum_k\beta_k^*\beta_k e^{i s \varphi}\right)+   \nonumber \\
  & f(-\pi+\varphi)\left(\sum_l\beta_l^*\alpha_l e^{i s (-\pi+\varphi)}\sum_m\alpha_m^*\beta_m e^{i s (-\pi+\varphi)}\right) \Biggr| ^2,
\end{align}
which reduces to (\ref{eqn:stdSSC}).
However, it produces atomic energy level shifts discussed in the next section. In any case, this is a peculiar interaction and appears unrealised in nature.

\section{Comparison with electron in orbit}
\label{sec:orbit}

\paragraph*{}
The crucial assumption made in \cite{unni2004} is that the phase obtained using $H_{int}=\bm{s}.\bm{B}$ in the rotating CM frame in which particles forward-scatter gives the phase accrued in the non-rotating CM frame.

\paragraph*{}
Now, would such a spin-gravito-magnetic phase accrue on a single particle scattering off a fixed potential and further also on a particle in orbit say in a storage ring or on an electron in a Bohr atom orbit? And in which frame should $H_{int}=\bm{s}.\bm{B}$, if at all, be applied in these cases? Indeed the spirit seems to be that a change in momentum direction is equivalent to universe rotating. Thus, one must take a co-rotating\footnote{Here, co-rotating means rotating at the rate at which the linear momentum changes direction. This co-rotating frame need not be co-moving. Thus it could be a frame with origin at nucleus and rotating at angular velocity of electron} frame and apply an interaction of the form $H_{int}=\bm{s}.\bm{B}$ in it to yield the lab frame phase. Or else, this prescription of a co-rotating frame applies only for particle scattering.

\paragraph*{}
Applying this prescription to an electron in atomic orbit we should obtain the gravito-magnetic phase on it in the lab frame. $B_z=-\omega$ here, where $\omega$ is the angular velocity of the electron. Such a spin-dependent phase due to gravito-magnetic interaction on the orbiting atomic electron would manifest as a spin-dependent energy splitting between spin-up and spin-down states. There are caveats here of picturing an electron in atomic orbit and assuming similar effects for bound vs scattering states. Further, it is possible that since the electron is revolving rather than merely rotating, there are other spin-dependent effects that cancel the spin-gravito-magnetic effect.

\paragraph*{}
The only relevant splitting observed experimentally that depends on the spin and the angular-velocity is the fine structure splitting due to spin-orbit coupling. It has two parts, the first due to the electric field of the nucleus in which gravity has no role and the second, Thomas correction (related to Thomas precession), is of magnitude $-s(1-\gamma)\omega$, where $\gamma=\frac{1}{\sqrt{1-v^2}}$ and $v$ the speed of the electron (Thomas correction is approximately half that of the first and opposite in sign). It is the second part which, if at all, may come from gravity. In non-relativistic case ($\gamma\approx 1$), this term is too small to allow the spin-gravito-magnetic effect $-s\omega$ as per the above prescription.

\paragraph*{}
Further, this prescription would also cause different shifts for different azimuthal quantum numbers $l$. Note that the interaction that works for all spin states namely (\ref{eqn:works}) would have the z-axis defined perpendicular to plane of orbit, and would also cause different shifts for different azimuthal quantum numbers $l$. 

\section{Rotation based proofs}
\label{sec:rotation}

\paragraph*{}
A typical rotation based proof of SSC \cite{broyles1976} postulates that physical exchange of identical particles into the same configuration must leave the total wavefunction unchanged (not even a phase change). There, physical exchange operators $R_{ab}$ are constructed for every pair $(a,b)$ of identical particles, which rotate only those two particles by $\pi$ about specially chosen axes. $R_{ab}$ actually rotates different simultaneous spin eigenkets $\ket{m_s^a,m_s^b}$ of the particle pair $(a,b)$ by $\pi$ around different axes, such that particles $a$ and $b$ end up exchanged back into their initial spatial and spin configuration.

\paragraph*{}
Thus from the action of $R_{ab}$ 
\begin{equation}
R_{ab}\psi(a,b,\ldots)=e^{i 2\pi s}\psi(b,a,\ldots),
\end{equation}
then using the postulate of invariance of wavefunction under $R_{ab}$, we get symmetrization/anti-symmetrization for bosons/fermions
\begin{equation}
\psi(a,b,\ldots)=e^{i 2\pi s}\psi(b,a,\ldots)=\pm\psi(b,a,\ldots).
\end{equation}

\paragraph*{}
Both the gravito-magnetic approach and the rotation approach involve some kind of rotation to generate a spinor phase. However the rotation approach works for arbitrary spins while the dynamical one does not (shown in section \ref{sec:arbitrary}). This is because the dynamical approach involves rotation about a fixed axis perpendicular to the plane of scattering, irrespective of the particles' initial spin states. But the rotation approach uses a physical exchange operator that rotates different simultaneous spin eigenkets $\ket{m_s^a,m_s^b}$ of a pair $(a,b)$ around different axes.

\paragraph*{}
Consider the scattering example in subsection \ref{subsec:example}, i.e. both spins along $x$ axis and axes as in figure \ref{fig:arbitrary_qmech_scattering}. The joint spin state of both particles is (in terms of $\ket{s_z}$ basis for each spin, $\ket{\downarrow}\equiv\ket{s_z=-\frac{1}{2}}$ and $\ket{\uparrow}\equiv\ket{s_z=\frac{1}{2}}$):
\begin{equation}
\frac{1}{\sqrt{2}} (\ket{\downarrow}+\ket{\uparrow}) 
\otimes \frac{1}{\sqrt{2}}(\ket{\downarrow}+\ket{\uparrow})
=\frac{1}{2} (\ket{\downarrow\downarrow}+\ket{\downarrow\uparrow}+\ket{\uparrow\downarrow}+\ket{\uparrow\uparrow}).
\end{equation}
The dynamical approach would use a frame rotating about the $z$-axis. However, the physical exchange operator $R_{ab}$ would rotate $\ket{\downarrow\downarrow}$ and $\ket{\uparrow\uparrow}$ by $\pi$ about the $z$ axis, but would rotate $\ket{\downarrow\uparrow}$ and $\ket{\uparrow\downarrow}$ by $\pi$ about the $y$ axis, to obtain the same spin configuration after physical exchange.

\paragraph*{}
Thus the rotation based proof uses contrived physical exchange operators and escapes the objections to the similar-seeming dynamical proposal. Of course the rotation based proof's basic premise of postulating a certain behaviour of the wavefunction (instead of observables) under transformations has been elsewhere questioned \cite{hilborn1995}.

\section{Conclusion}
\label{sec:conclusion}
The proposal of \cite{unni2004} to obtain the SSC via spin-cosmic-gravity dynamics does not reproduce the scattering cross-sections given by the standard implementation of SSC in quantum mechanics, for most initial spin states of scattering identical particles. Further, it could cause (with caveats) a large spin-orbit effect in atomic energy levels which is not measured. No simple dynamical interaction seems available to account for the SSC.


\section{Acknowledgements}
I thank Prof. C S Unnikrishnan for introducing me to this problem and for extensive and patient discussions on the same. I also thank Rakesh Tibrewala and R. Loganayagam for perusal and insightful comments on the article.


\begin{thebibliography}{}
%
%

\bibitem{unni2004}
C S Unnikrishnan, Spin-statistics connection and the gravity of the universe: the cosmic connection, \textit{Preprint} gr-qc/0406043 (2004).

\bibitem{duck1997}
Ian Duck and E C G Sudarshan, Pauli and the spin-statistics theorem. World Scientific, Singapore (1997).

\bibitem{unni2004a}
C S Unnikrishnan, Cosmic relativity: the fundamental theory of relativity, its implications, and experimental tests, \textit{Preprint} gr-qc/0406023 (2004).

\bibitem{unni2008}
C S Unnikrishnan, private communication, early 2008.

\bibitem{mashhoon1988}
Bahram Mashhoon, Neutron interferometry in a rotating frame of reference, Phys. Rev. Lett., 61, pp. 2639-2642 (1988).

\bibitem{tino2000}
Guglielmo M Tino, Testing the symmetrization postulate of quantum mechanics and the spin-statistics connection, Fortschr. Phys., 48, pp. 537-543 (2000).

\bibitem{kessler1976}
Joachim Kessler, Polarized electrons, pp. 87--91. Springer-Verlag, Berlin Heidelberg (1976).

\bibitem{taylor1972}
John R Taylor, Scattering theory: the quantum theory on nonrelativistic collisions, pp. 453--449. John Wiley \& Sons, Inc., New York (1972).

\bibitem{broyles1976}
A A Broyles, Spin and statistics, American Journal of Physics, 44, pp. 340--343 (1976).



\bibitem{hilborn1995}
R C Hilborn, Questions and answers, American Journal of Physics, 63, pp. 298--299 (1995).

\end{thebibliography}


\end{document}